\documentclass[a4paper,fleqn,usenatbib]{mnras}

\usepackage[T1]{fontenc}
\usepackage{ae,aecompl}

\usepackage[normalem]{ulem}
\usepackage{amsmath}
\usepackage{graphicx} 
\usepackage{lscape}
\usepackage{indentfirst}
\usepackage{enumitem}

\usepackage{amssymb}
\usepackage{color}
\usepackage[flushleft]{threeparttable}
\newcommand {\bc}{\begin {center}}
\newcommand {\ec}{\end {center}}
\newcommand {\be}{\begin {equation}}
\newcommand {\ee}{\end {equation}}
\newcommand {\beq}{\begin {eqnarray}}
\newcommand {\eeq}{\end {eqnarray}}
\newcommand {\comment}[1]{}

\newcommand{\red}[1]{\textcolor{black}{#1}}

\renewcommand{\d}{{\rm d}}

\newcommand {\ergs}{{\rm erg\ \rm s^{-1}}}

\title[Outflows and beaming in bright XRPs]
{Bright X-ray Pulsars: how outflows influence beaming, pulsations and pulse phase lags}
\author[A. A.~Mushtukov and S. Portegies Zwart] 
{Alexander~A.~Mushtukov$^{1,2}$\thanks{E-mail: alexander.mushtukov@physics.ox.ac.uk (AAM)}
and
Simon Portegies Zwart$^{2}$
\\ 
$^1$ Astrophysics, Department of Physics, University of Oxford, Denys Wilkinson Building, Keble Road, Oxford OX1 3RH, UK\\
$^2$ Leiden Observatory, Leiden University, NL-2300RA Leiden, The Netherlands \\
} 

\pubyear{2022}

\begin{document}
\label{firstpage}
\pagerange{\pageref{firstpage}--\pageref{lastpage}}
\maketitle

%%%%%%%%%%%%%%%%%%%%%%%%%%%%%%%%%%%%%%%%%%%%%%%%%%

\begin{abstract} Extreme accretion in X-ray pulsars (XRPs) results in
radiation-driven outflows launched from the inner parts of the
accretion disc.  The outflows affect the apparent luminosity of the
XRPs and their pulsations through the geometrical beaming.  We model
processes of geometrical beaming and pulse formation using Monte Carlo
simulations.  We confirm our earlier statement that strong
amplification of luminosity due to the collimation of X-ray photons is
inconsistent with a large pulsed fraction.  Accounting for
relativistic aberration due to possibly high outflow velocity ($\sim
0.2c$) does not affect this conclusion.  We demonstrate that the
beaming causes phase lags of pulsations.  Within the opening angle of
the accretion cavity formed by the outflows, phase lags tend to be
sensitive to observers viewing angles.  Variations in outflow geometry
and corresponding changes of the phase lags might influence the
detectability of pulsation in bright X-ray pulsars and ULXs.  We
speculate that the strong geometrical beaming is associated with large
radiation pressure on the walls of accretion cavity due to multiple
photons reflections.  We expect that the mass loss rate limits
geometrical beaming: strong beaming becomes possible only under
sufficiently large fractional mass loss rate from the disc.
\end{abstract}


%%%%%%%%%%%%%%%%%%%%%%%%%%%%%%%%%%%%%%%%%%%%%%%%%
\section{Introduction}
\label{sec:Intro}
%%%%%%%%%%%%%%%%%%%%%%%%%%%%%%%%%%%%%%%%%%%%%%%%%

X-ray pulsars (XRPs) are accreting strongly magnetized neutron stars (NSs) in close binary systems (see \citealt{2022arXiv220414185M} for review).
Typical magnetic field strength at the NS surface in XRPs is expected to be $B\sim 10^{12}\,{\rm G}$.
The apparent luminosity of XRPs covers many orders of magnitude from $\sim 10^{32}\,\ergs$ up to $\sim 10^{41}\,\ergs$.
The brightest XRPs belong to the class of recently discovered pulsating ultraluminous X-ray sources (ULXs, see \citealt{2014Natur.514..202B,2017Sci...355..817I} and \citealt{2021AstBu..76....6F} for review) and Be-X-ray transients at the peak of their outbursts \citep{2017A&A...605A..39T,2019ApJ...873...19T,2020MNRAS.491.1857D,2020MNRAS.494.5350V,2020MNRAS.495.2664C}.
The geometry of the accretion flow in XRPs is strongly affected by the magnetic field of a NS.
The accretion flow from the companion star in the form of wind or accretion disc is destroyed at a certain distance called the magnetospheric radius, where magnetic field pressure starts dominating the ram pressure of accreting material. 
The radius of NS magnetosphere in XRP depends on the mass accretion rate $\dot{M}$ and NS magnetic field strength and structure. 
For the case of magnetic field dominated by the dipole component, the magnetospheric radius is estimated as 
\beq\label{eq:R_A}
R_{\rm m}
= 1.8\times 10^8\,\Lambda B_{12}^{4/7}\dot{M}_{17}^{-2/7}m^{-1/7}R_6^{12/7}\,\,{\rm cm},
\eeq
where $\Lambda\sim 0.5 - 1$ is a factor depending on the accretion flow geometry and physical condition (see, e.g., \citealt{2019A&A...626A..18C}), $B_{12}$ is a surface magnetic field strength in units of $10^{12}\,{\rm G}$, 
$\dot{M}_{17}$ is mass accretion rate in units of $10^{17}\,{\rm g\,s^{-1}}$, 
$m$ is a mass of a NS in units of $M_\odot$, and $R_6$ is the NS radius in units of $10^6\,{\rm cm}$.
When observing from the magnetospheric radius, the accretion flow follows magnetic field lines and reaches the NS surface in a small region close to a NS's magnetic poles.
If the inner disc radius becomes small enough at high mass accretion rates:
\beq
R_{\rm m} < R_{\rm A}\approx 2.7\times 10^{8}\,\dot{M}_{19}^{16/21}m^{7/21}\,{\rm cm},
\eeq 
the inner parts of accretion disc become radiation pressure dominated \citep{1973A&A....24..337S}.
At 
\beq\label{eq:dotM_lim_d}
\dot{M} > 5\times 10^{19} \Lambda^{7/9} B_{12}^{4/9} m^{2/3} R_6^{4/3}\,\,{\rm g\,s^{-1}},
\eeq
accretion disc starts to lose material in the form of radiation-driven outflows.

Non-dipole magnetic fields
\footnote{The evidence of non-dipole magnetic field structure was reported already in a few XRPs including Her~X-1 \citep{2022MNRAS.515..571M}, NGC~5907~X-1 \citep{2017Sci...355..817I} and Galactic Be transient Swift~J0243.6+6124 \citep{2022ApJ...933L...3K}.}
decreases faster with the distance from a NS, and the accretion disc can come closer \red{(see numerical simulations performed by \citealt{2007MNRAS.374..436L,2008MNRAS.386.1274L,2022MNRAS.515.3144D})}.
In the case of a magnetic field dominated by a quadrupole component, the magnetospheric radius can be estimated as
\beq\label{eq:R_A_quad}
R_{\rm m}^{\rm (q)}\sim
3.44\times 10^{7}\,
B_{12}^{4/11}\dot{M}_{17}^{-2/11}m^{1/11}R_6^{16/11} \,{\rm cm},
\eeq 
which results in the launch of the outflows at slightly lower mass accretion rates:
\beq\label{eq:Mdot_lim_q}
\dot{M}^{\rm (q)}\gtrsim 2\times 10^{19}\,B_{12}^{4/13} m^{12/13} R_6^{16/13}\,\,{\rm g\,s^{-1}}.
\eeq
According to (\ref{eq:dotM_lim_d}) and (\ref{eq:Mdot_lim_q}), considerable mass loss rates due to radiation driven outflows becomes possible only for sufficiently high mass accretion rates from a companion star or/and sufficiently weak magnetic field of a NS (see, e.g., \citealt{2019MNRAS.484..687M,2019A&A...626A..18C}).
The evidence of the outflows in a few bright X-ray binaries, including pulsating ULX NGC~300~X-1 \citep{2018MNRAS.479.3978K}, and bright Galactic BeXRP Swift~J0243.6+6124 \citep{2019MNRAS.487.4355V} as ware already reported.
In both cases, the velocity of the outflows was measured to be $v\sim 0.2 c$.

%%%%%%%%%%%%%%%%%%%%%%%%%%%%%%%%%%%%%%%
\begin{figure*}
\centering 
\includegraphics[width=15.cm]{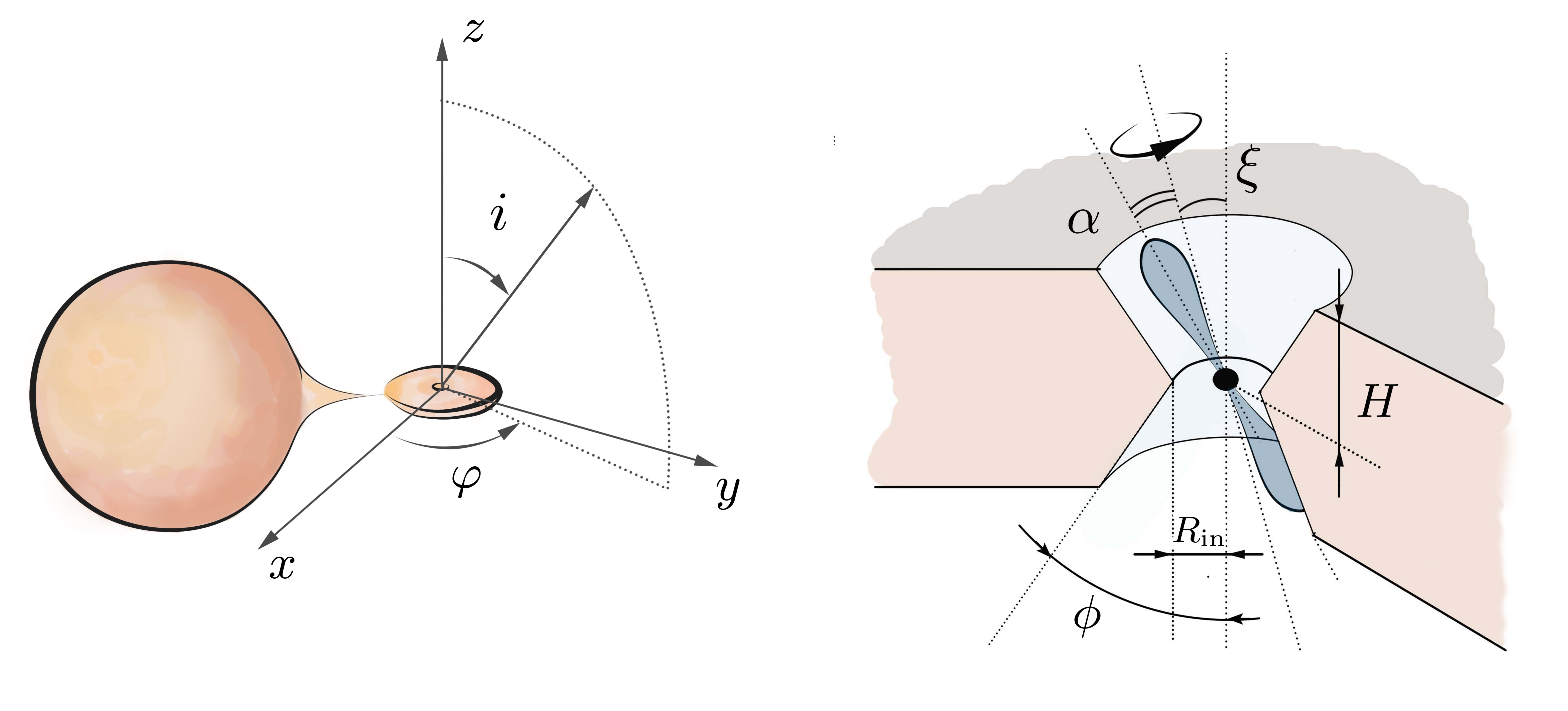}
\caption{
Schematic illustration of the considered geometry. 
Under conditions of extremely intensive accretion in a binary system, the accretion flow (accretion disc and outflow from the disc) from the companion star forms an accretion cavity around the central object.
The cavity naturally collimates X-ray flux from a NS and affects the pulsation properties. 
}
\label{pic:scheme}
\end{figure*}
%%%%%%%%%%%%%%%%%%%%%%%%%%%%%%%%%%%%%%%

The apparent luminosity $L_{\rm app}$ of the XRP is obtained from their X-ray energy flux averaged over the pulse period $F_{\rm ave}$:
$L_{\rm app}\approx 4\pi F_{\rm ave}D^2$, where $D$ is a distance to the XRP.
The outflows 
\red{place NS into the accretion cavity and} 
collimate radiation from the central object, making the apparent luminosity of a source smaller or larger than the actual luminosity, depending on the mutual orientation of a system and distant observer \citep{2009MNRAS.393L..41K,2017MNRAS.468L..59K,2021ApJ...917L..31A}.
The coefficient of proportionality $a$ between the apparent and actual luminosity is the amplification factor:
\beq 
L_{\rm app}=a L.
\eeq 
The apparent luminosity and amplification factor depend on the geometry of accretion flow/outflow, and varies for distant observers.

\red{The geometrical collimation and corresponding multiple reflections of X-ray photons from the walls of the accretion cavity influence pulsations in bright XRPs  \citep{2021MNRAS.501.2424M}.}
In particular, the strong geometrical beaming of X-ray radiation lead to a considerable reduction in the pulsed fraction (PF)
\beq
{\rm PF}=\frac{F_{\rm max}-F_{\rm min}}{F_{\rm max}+F_{\rm min}},
\eeq
where $F_{\rm min}$ and $F_{\rm max}$ are minimal and maximal fluxes during the pulsation period.
The possibly large velocity of the outflow can influence reprocessing of X-ray photons by the walls of the accretion cavity: relativistic aberration can reduce the number of reflections and therefore increase the pulse fraction.
In addition, reflections of the beam pattern from the walls of the accretion cavity can change the pulse shape, and cause phase lags in observed pulsations. 

This paper follows our recent analyses of geometrical beaming
performed in \citep{2021MNRAS.501.2424M}.  Using Monte Carlo
simulations, we investigate more realistic geometry of
radiation-driven outflows from accretion disc (see
Fig.\,\ref{pic:scheme}) and account for relativistic aberration due to
the possibly high velocity of the outflows.  We analyze the influence
of geometrical beaming and outflow velocity on the PF
(Section\,\ref{sec:PF_influenced_by_beaming}) and phase shift of X-ray
pulsation (Section\,\ref{sec:Phase_shift}).  The repeated reprocessing
of X-ray photons inside the accretion cavity influences the radiation
pressure and therewith the geometry of the accretion cavity
(Section\,\ref{sec:RadPressure}).

%%%%%%%%%%%%%%%%%%%%%%%%%%%%%%%%%%%%%%%%%%%%
\section{Model}
\label{sec:Model}
%%%%%%%%%%%%%%%%%%%%%%%%%%%%%%%%%%%%%%%%%%%%

\red{Geometry of radiation-driven outflows in XRPs is determined by multiple factors, including the structure of NS magnetosphere, details of wind launch and mechanisms of opacity.}
We consider \red{a simplified} geometry where the NS is located inside the cavity of a fixed geometrical thickness. 
The walls of a cavity are given by two conical surfaces with opening angle $\phi$ (see Fig.\,\ref{pic:scheme}\, right).
The geometry is therefore determined by three parameters: the inner radius of the accretion flow $R_{\rm in}$, the effective semi-thickness of the cavity $H$, and the opening angle of the conical surfaces $\phi$.
Two angles determine NS rotation: 
The angle between the orbital axis and the NS spin axis is $\xi$, and the magnetic obliquity is $\alpha$, i.e., the angle between the rotational axis and the magnetic axis of the NS.
The $z$-axis is aligned with the rotational axis of the accretion flow, the NS spin is in the $x-z$ plane, and the $y$-axis completes the right-handed coordinate system. 
The observer's position is specified by the inclination $i$ and azimuthal angle $\varphi$ (see Fig.\,\ref{pic:scheme}\, left).
We assume that most of the accretion energy is emitted in close proximity to the magnetic poles of a NS.
Rotation of a NS results in a specific average flux and pulsations detectable by a distant observer. 
The average flux and pulsed signal depend on specific rotation parameters, the geometry of accretion flow, and the position of the observer with respect to the system. 

Using Monte Carlo simulations, we trace the history of photons emitted near the NS surface and can reproduce the shape of pulsation in any given direction.
Our code is designed under the following set of assumptions: 
\begin{itemize}[leftmargin=15pt]
\item A dipole component dominates the magnetic field of a NS, and an accreting NS produces two beams directed along the magnetic axis.
The angular distribution of the luminosity is assumed to follow:
\beq\label{eq:beam}
\frac{\d L(\theta)}{\d\cos\theta}\propto \cos^n\theta,
\eeq
where $\theta\in[0;\pi/2]$ is the angle between the magnetic axis and the photon momentum. 
\item Reflection of a photon by the walls of the accretion funnel is calculated under the assumption of multiple conservative, isotropic (in the reference frame co-moving with the outflow), and coherent scattering in the semi-infinite medium.
\item The material that forms the walls of the accretion cavity can have finite velocity. 
The velocity vector at the edge of the accretion cavity is assumed to be along the conical surface and is directed away from the plane of accretion flow symmetry.
\item The light travel time inside the accretion funnel is assumed to be much shorter than the NS spin period and thus does not affect the formation of the pulse profile  (applicability of this assumption is discussed in \citealt{2021MNRAS.501.2424M}, Section 3.3). 
\end{itemize}

To account for non-zero outflow velocity, we adopt that the walls of the accretion cavity move with a constant velocity $v$ in the direction opposite to the central object. 
Due to the relativistic aberration, the momentum direction of a photon that reaches the edges of the accretion cavity is different in the laboratory reference frame and the reference frame co-moving with the outflow material.
Simulation of the reflection process consists of three steps:
(i) the direction of photon motion is recalculated from the laboratory reference frame to the reference frame co-moving with the outflow, 
(ii) using the Monte-Carlo approach and assuming a semi-infinite medium, we calculate the direction of a reflected photon in the co-moving reference frame, and 
(iii) the direction of photon momentum is recalculated from the co-moving reference frame to the laboratory reference frame.
We use Lorentz transformation to recalculate the direction of photon momentum from one reference frame to another. 
Defining the local coordinate system with the $z'$-axis along the direction of motion of the gas and the $x'$-axis along the normal to the surface, we can write the unit vector along the photon momentum in the laboratory reference frame as 
\beq 
\bf{n}_\omega = (\sin\chi\cos\psi,\sin\chi\sin\psi,\cos\chi).
\eeq
The unit vector along the photon momentum in the reference frame co-moving with the gas is given by \citep{2013ApJ...777..115P}
\beq\label{eq:n_}
{\bf n}_\omega' 
&=& (\sin\chi'\cos\psi',\sin\chi'\sin\psi',\cos\chi') \nonumber \\
&=&
\left(D\sin\chi\cos\psi,D\sin\chi\sin\psi,\frac{\cos\chi-\beta}{1-\beta\cos\chi}\right),
\eeq 
where
\beq\label{eq:D_gamma}
D=\gamma^{-1}(1-\beta\cos\chi)^{-1},\quad 
\gamma=(1-\beta^2)^{-1/2}.
\eeq 
The inverse transformation from the co-moving reference frame to the laboratory one can be obtained from (\ref{eq:n_}) and (\ref{eq:D_gamma}) by replacing $\beta$ with $(-\beta)$.

Tracing a history of $N$ photons emitted by a NS at specific geometrical configurations ($N=10^9$ in the simulations represented in this paper), we can reproduce the pulse profiles as different distant observers observe them. 
On the base of constructed pulse profiles, we obtain the average flux and, therefore, the luminosity amplification factor $a$ (see details in \citealt{2021MNRAS.501.2424M}).
Note that some difference between actual and apparent luminosity appears even without the accretion cavity collimating X-ray photons. 
The constructed pulse profiles naturally provide us with the PF. 
Cross-correlating the pulse profiles constructed for the accretion flow absence with the pulse profiles calculated for a given set of accretion cavity parameters, we can get the phase lag of X-ray pulsations caused by geometrical beaming.
Similarly to the luminosity amplification factor and PF, the phase lags depend on the mutual orientation of the binary system and the distant observer.
Assuming the random mutual orientation of a system in the observer's reference frame, we construct the expected distributions of accreting NSs over the amplification factors and obtain the average PF and the average phase lag of the pulsations for each value of $a$.

The photons experiencing geometrical beaming in the accretion cavity participate in many reflection events at the edge of the cavity. 
Each reflection event is accompanied by a transfer of photon momentum to the walls of the cavity. 
One would expect that in the case of stronger geometrical beaming, the momentum transfer to the walls of the cavity, i.e., to the outflow, is larger because the photons experience more reflections inside the cavity before leaving the system. 
If the momentum transferred to the walls of the accretion cavity is comparable to the total momentum of the mass outflow rate in XRP, The radiation pressure inside the accretion cavity is high enough to affect the geometry of the outflow.
We do not model the influence of radiation pressure on the outflow geometry but estimate the total photon momentum transfer to the walls of the accretion cavity.

%%%%%%%%%%%%%%%%%%%%%%%%%%%%%%%%%%%%%%%%
\section{Numerical results}
\label{sec:NumResults}
%%%%%%%%%%%%%%%%%%%%%%%%%%%%%%%%%%%%%%%%

%----------------------------------------%
\begin{figure}
\centering 
\includegraphics[width=8.5cm]{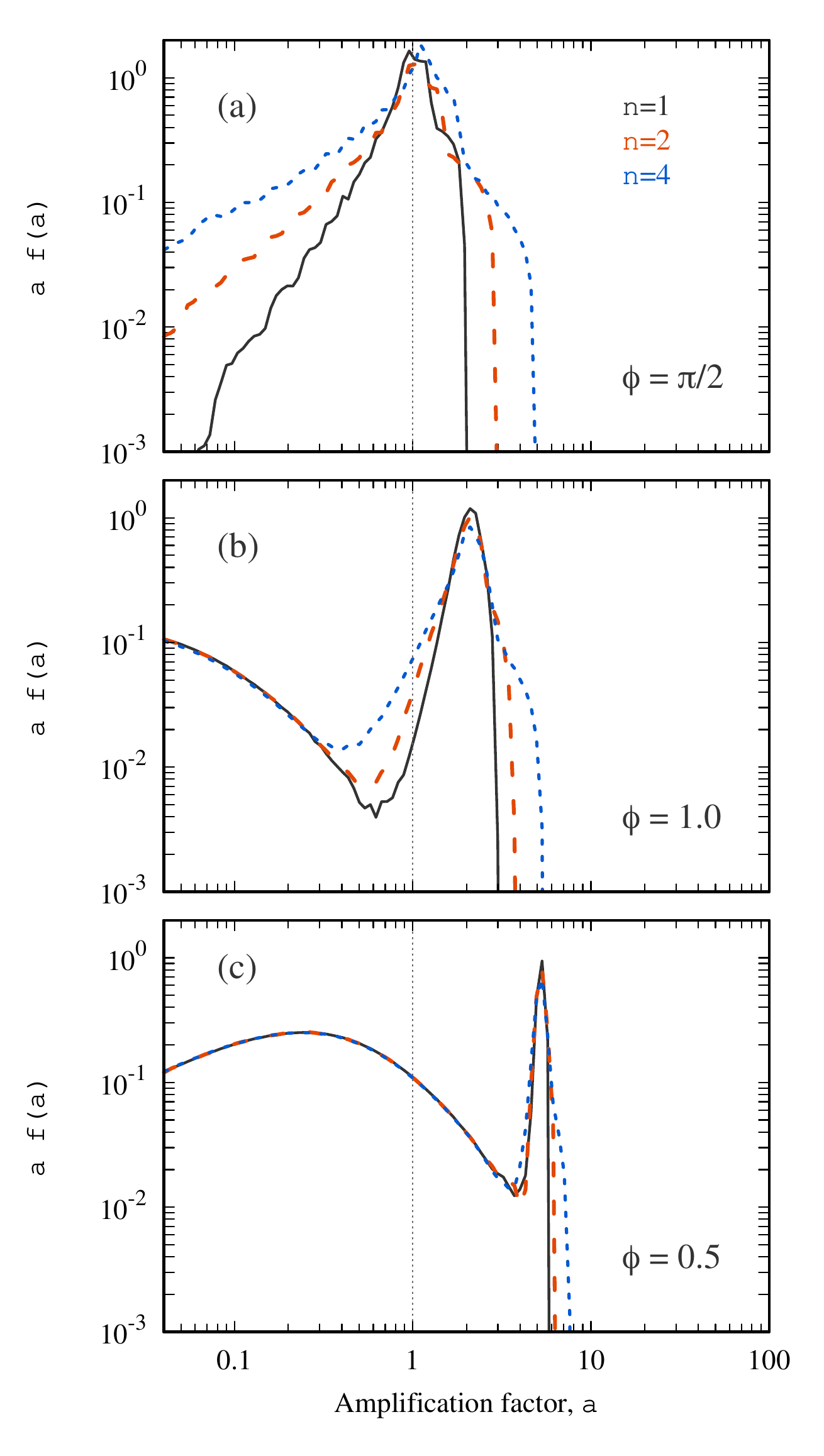} 
\caption{
The differential distributions of sources over the amplification factor. 
Different lines represent distributions corresponding to different parameters $n$: $1$ (black solid), $2$ (red dashed) and $4$ (blue dotted).
Different panels correspond to different opening angles of accreting cavity: (a) $\phi=\pi/2$, (b) $\phi=1$, and (c) $\phi=0.5$.
One can see that even in the case of the absence of geometrical beaming  ($\phi=\pi/2$, panel a), the apparent luminosity can differ from the actual one, and the distribution of sources over the amplification factor is affected by the initial beam pattern.
On the contrary, the initial beam pattern of radiation does not affect much the distribution of sources over the amplification factor in the case of a sufficiently small opening angle $\phi$ (see panels b and c).
Parameters: $H/R_{\rm in}=100$.
}
\label{pic:sc_fL}
\end{figure}
%-----------------------------------------%

%----------------------------------------%
\begin{figure*}
\centering 
\includegraphics[width=17.5cm]{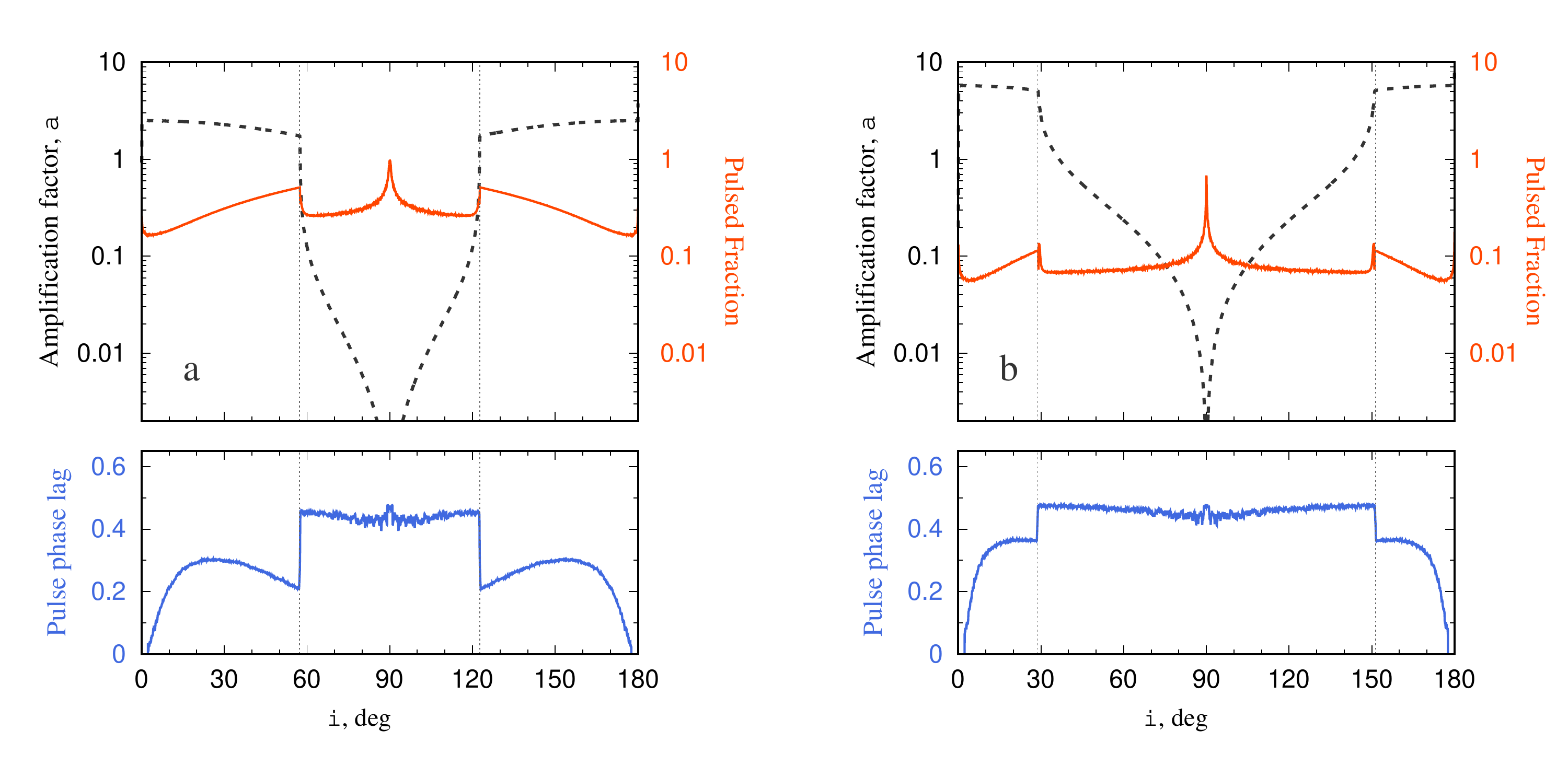} 
\caption{
The dependence of averaged amplification factor (black dashed lines), pulsed fraction (solid red lines) and pulse phase lags(solid blue lines on the bottom panel) on the observer's inclination $i$ with respect to the accretion flow plane. 
The left and right panels are given for different opening angles of the accretion cavity: $\phi=1\,{\rm rad}$ and $\phi=0.5\,{\rm rad}$ respectively.
Vertical dotted lines give the opening angles.
Parameters: $H/R_{\rm in}=100$, $n=2$.
}
\label{pic:amp_and_PF}
\end{figure*}
%-----------------------------------------%

The numerical results presented below are averaged over the parameters of NS rotation given by $\alpha$ and $\xi$ and displacement of a distant observer in respect to the system given by angles $i$ and $\varphi$ (see Fig.\,\ref{pic:scheme}).
The rotational axis of the NS is taken to be randomly oriented to the accretion flow.
The angle between the rotation axis and magnetic axis $\alpha$ takes random values in the interval $[0;\pi]$.
Note that the accretion process onto magnetized NS tends to align the rotational axis of a star with the accretion disc axis. 
The relaxation time to the equilibrium depends on the magnetic moment of a NS \citep{1982SvA....26...54L}. 
The rotational axis is expected to be aligned with the disc axis of the XRPs, which hosts a NS with an extremely strong magnetic field. At the same time, the orientation of the rotational axis of a weakly magnetized NS can be far from equilibrium and randomly oriented.
The results from aligning the rotational axis with the axis of the accretion disc are only slightly different than those obtained from a random orientation of the rotational axis.
The initial beam pattern is taken to be described using $n=2$ in our simulations (see eq.\,\ref{eq:beam}).
In the case of strong geometrical beaming, variations of $n$ affect the final results insignificantly (see Fig.\,\ref{pic:sc_fL}).

%--------------------------------------------%
\begin{figure*}
\centering 
\includegraphics[width=17cm]{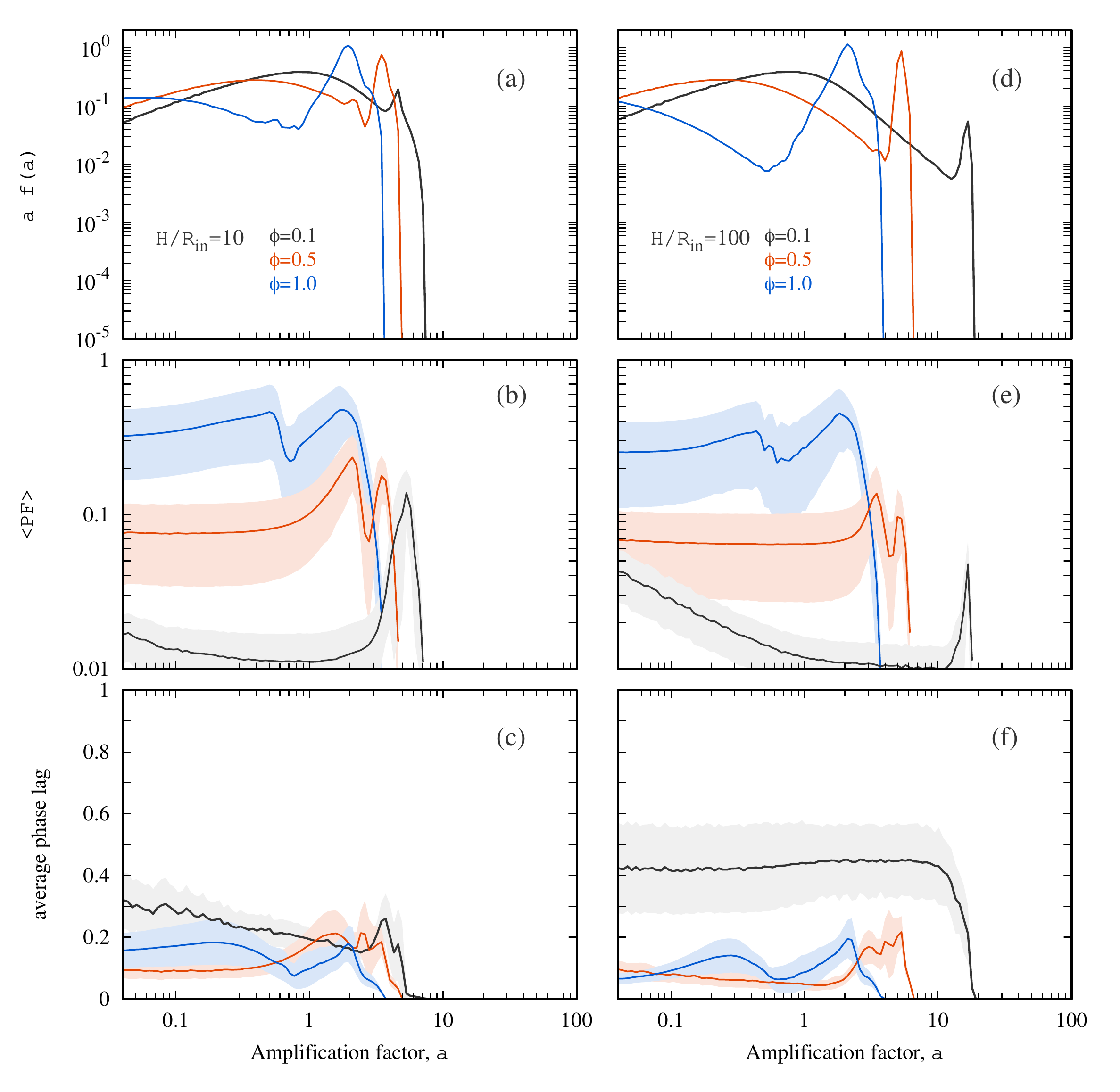} 
\caption{
The expected features of XRPs at different amplification factors $a$:
(i) The differential distribution over the amplification factor (a,d).
One can see that thicker accretion flow (i.e., larger $H/R_{\rm in}$) and smaller opening angle of accretion cavity $\phi$ result in larger amplification factors in the distribution.
(ii) The average PF as a function of the amplification factor (solid lines) and the regions within one standard deviation from the average value (b,e).
(iii) The average phase lags of pulsations at a given amplification factor (c,f). 
Different curves correspond to different opening angles of the accretion cavity surface: $\phi=0.1$ (black), $0.5$ (red) and $1$ (blue).
Left and right columns correspond to $H/R_{\rm in}=10$ and $H/R_{\rm in}=100$ respectively.
}
\label{pic:sc_beaming}
\end{figure*}
%--------------------------------------------%

%%%%%%%%%%%%%%%%%%%%%%%%%%%%%%%%%%%%%%%%%%%%%%%%%
\subsection{Geometrical beaming influencing pulsed fraction}
\label{sec:PF_influenced_by_beaming}
%%%%%%%%%%%%%%%%%%%%%%%%%%%%%%%%%%%%%%%%%%%%%%%%%

Strong geometrical beaming in the directions close to the axis of symmetry is achievable at large $H/R_{\rm in}$ ratios and small opening angles of accretion cavity $\phi$ (see Fig.\,\ref{pic:amp_and_PF}).
Outside the opening angle of the cavity, the luminosity amplification factor drops rapidly and turns to zero at $i=\pi/2$. 
The beaming results in a drop of the averaged pulse fraction (see Fig.\,\ref{pic:sc_beaming}a,b and d,e).
The PF above $10$ per cent is hardly possible for amplification factors $a>10$, which is consistent with our earlier report \citep{2021MNRAS.501.2424M}.
The strongest pulsations are expected for observers with inclination $i\sim \phi$ with respect to the accretion disc plane (see Fig.\,\ref{pic:amp_and_PF}), but PF is still small in the case of a large amplification factor.
For the viewing angles within the opening angle of the accretion cavity, the PF can be variable within a factor of a few.
Large PF at $i\sim \pi/2$ is of no interest because of the very low apparent luminosity in this direction.

%%%%%%%%%%%%%%%%%%%%%%%%%%%%%%%%%%%%%%%%%%%%%%%%%
\subsection{Pulse phase shift due to the beaming}
\label{sec:Phase_shift}
%%%%%%%%%%%%%%%%%%%%%%%%%%%%%%%%%%%%%%%%%%%%%%%%%

Numerical simulations reveal the appearance of phase shift of X-ray pulsations due to the geometrical beaming (see lower panels in Fig.\,\ref{pic:amp_and_PF} and Fig.\,\ref{pic:sc_beaming}\,c,f).
This result seems to be natural because, in the case of beamed flux, the observer detects a significant fraction of reprocessed/reflected photons.
The phase shift depends on the geometry of the outflows and the observer's inclination with respect to the accretion flow plane (see the lower panel in Fig.\,\ref{pic:amp_and_PF}).
The lags are minimal for observers looking at the system along the axis of symmetry and strongly dependent on the observer's inclination $i$ even within the opening angle of the accretion cavity ($i<\phi$), where the amplification factor shows only slight variations.
Thus, the phase lags are more sensitive to the appearance of the outflow and variations of its geometry than the amplification factor.

In the case of strong geometrical beaming, the average phase lag is close to $0.5$ expect the largest amplification factors, which correspond to the case of distant observers looking at the system close to the axis of symmetry.
The dispersion measure of the lags tends to increase with the increase of the effective thickness of the flow and the decrease of the cone opening angle.

%%%%%%%%%%%%%%%%%%%%%%%%%%%%%%%%%%%%%%%%%%%%%%%
\subsection{Influence of the outflow velocity}
\label{sec:outflow_velocity}
%%%%%%%%%%%%%%%%%%%%%%%%%%%%%%%%%%%%%%%%%%%%%%%

%----------------------------------------%
\begin{figure}
\centering 
\includegraphics[width=8.5cm]{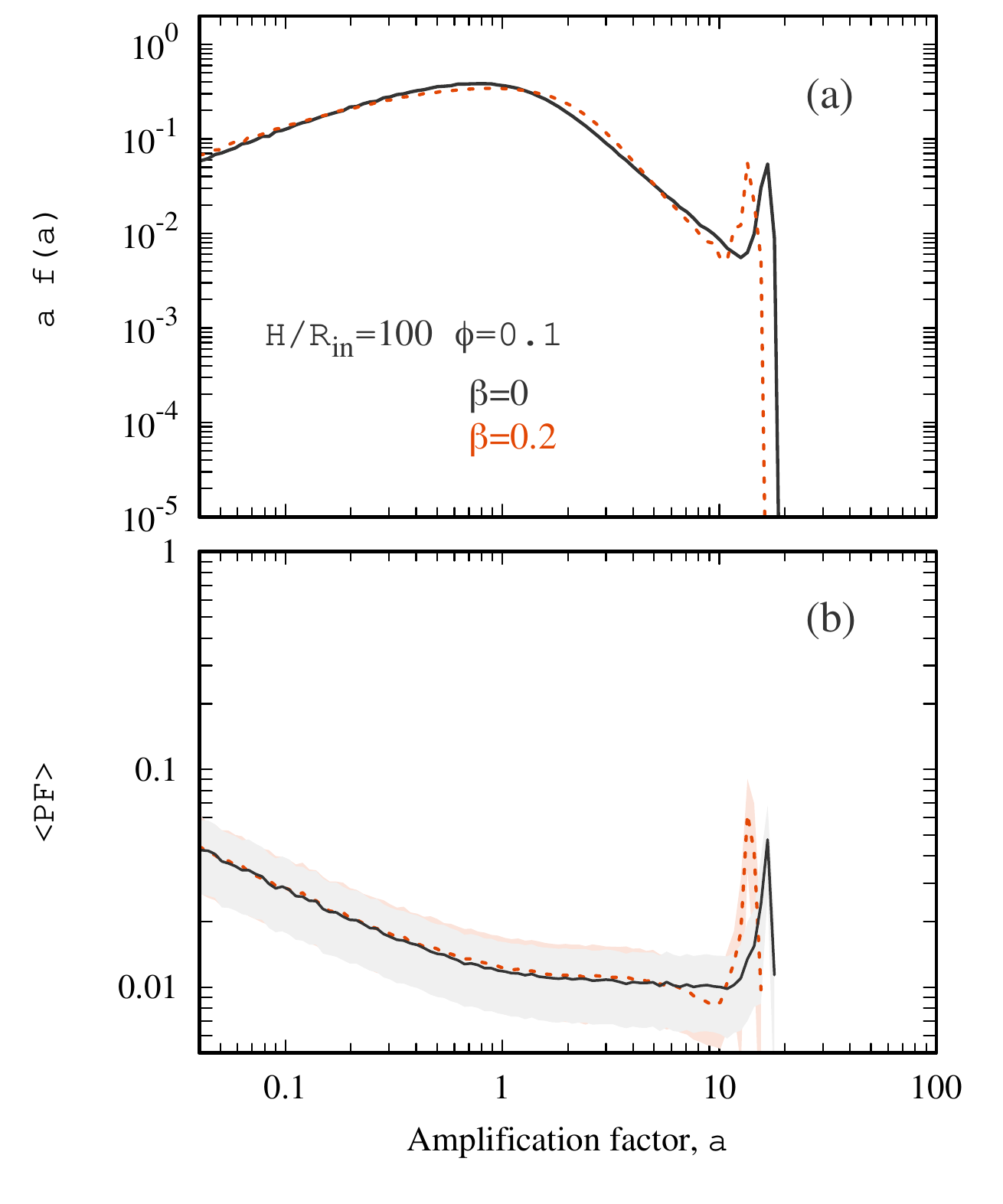} 
\caption{
The influence of outflow velocity $\beta$ on the distribution of sources over the amplification factor (a) and typical PF at a given amplification factor (b).
Black solid and red dashed lines are given for $\beta=0$ and $\beta=0.2$, respectively.
One can see that the influence of outflow velocity is negligible. 
}
\label{pic:sc_beaming_velosity}
\end{figure}
%-----------------------------------------%

We considered the outflow velocity $v=0.2c$ similar to the ones reported in bright X-ray transients and ULXs \citep{2018MNRAS.479.3978K,2019MNRAS.487.4355V,2021MNRAS.505.5058P}.
Our simulations accounting for relativistic aberration reveal only a slight influence of the outflow velocity on the sources distribution over the amplification factor, average PF, and phase lag (see Fig.\,\ref{pic:sc_beaming_velosity}).

%%%%%%%%%%%%%%%%%%%%%%%%%%%%%%%%%%%%%%%%%%%%%%%
\subsection{Radiation pressure on the walls of accretion cavity}
\label{sec:RadPressure}
%%%%%%%%%%%%%%%%%%%%%%%%%%%%%%%%%%%%%%%%%%%%%%%

\begin{figure}
\centering 
\includegraphics[width=8.8cm]{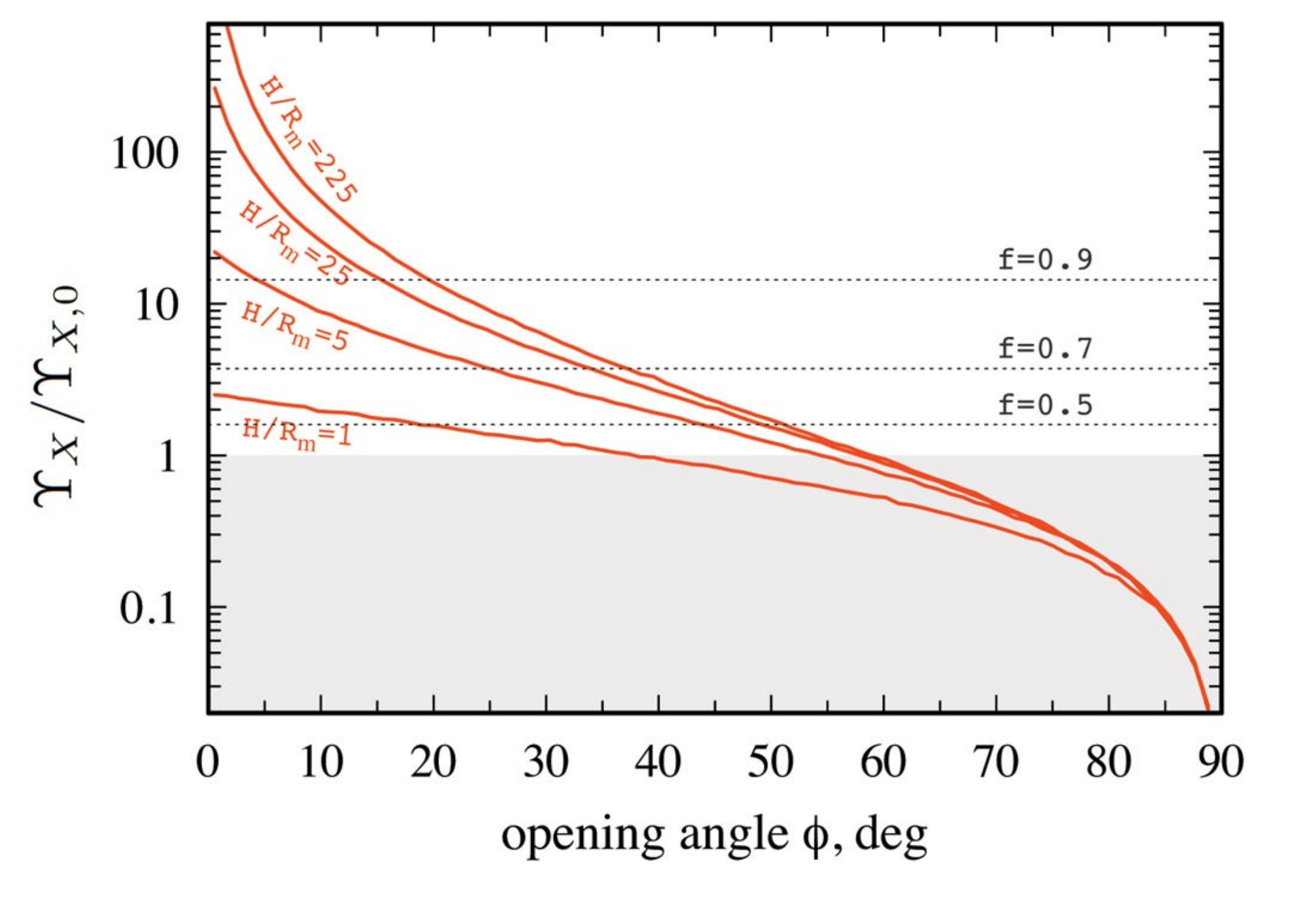} 
\caption{
The momentum transferring to the walls of the accretion cavity by X-ray photons (in units of the total momentum of the photons emitted at the NS surface) as a function of accretion cavity opening angle $\phi$.
Different curves are given for different relative thicknesses of the flow $H/R_{\rm in}=1,\,5,\,25,\,225$ (from bottom to top, respectively).
In the case of large $H/R_{\rm m}$ and small $\phi$, the photons experience multiple reflections and effectively transfer their momentum several times. It enforces radiation pressure on the walls.
Horizontal dotted lines (given for different fractions of material launched from the disc: $f=0.8,\,0.9,\,0.95$) show conditions when the momentum transferred to the outflow is comparable to its total momentum. 
Above a dotted line of corresponding $f$, radiation pressure is expected to be strong enough to affect the geometry of the outflow.
}
\label{pic:sc_Lp}
\end{figure}

The total momentum $L_{p}$ transfer (in a unit time) from X-ray photons locked inside the accretion cavity to the outflow in a direction perpendicular to the symmetry axis of the system is dependent both on the opening angle of the conical surfaces $\phi$ and effective geometrical thickness $H$ (see Fig.\,\ref{pic:sc_Lp}).
In the case of small opening angles $\phi$ and large $H/R_{\rm in}$ ratios, which are necessary conditions for a strong luminosity amplification, the transferred momentum is much greater than the total momentum of the photons emitted at the NS surface.

In the case of advective discs, the radiation-driven outflow hardly carries away more than half of accreting material even at a high mass accretion rate $\dot{M}_0$ from the companion star \citep{1999AstL...25..508L,2007MNRAS.377.1187P}.
Therefore, the total momentum carried by the wind per unit of time can be limited from above as 
\beq
\Upsilon_{\rm wind}\lesssim\frac{\dot{M}_0}{2}0.2c = \frac{\dot{M}_0 c}{10}.
\eeq
The total momentum carried by X-ray photons per unit of time at close proximity to the surface of accreting NS is
\beq
\Upsilon_{X,0}=\frac{L}{c}=\frac{GM\dot{M}}{Rc}\gtrsim
\frac{v_{\rm ff}^2}{c}\frac{\dot{M}_0}{2},
\eeq
where $M$ and $R$ are NS mass and radius, respectively, $\dot{M}\in[0.5\dot{M}_0;\dot{M}_0]$ is the mass accretion rate onto the NS surface, and $v_{\rm ff}$ is a free-fall velocity at NS surface.
Assuming $v_{\rm ff}\sim 0.5c$, one can estimate the momentum of X-ray photons from below as $L_{p,0}\gtrsim \dot{M}_0/8$.
Therefore, in the case of advective accretion discs, the momentum of photons emitted at the NS surface already exceeds the momentum of the wind launched from the disc.
Under this condition, strong geometrical beaming is impossible because the collimation of X-ray luminosity will result in radiation pressure on the walls of the accretion cavity strong enough to change the geometry of the outflows.

In the case of weak advection, a larger fraction of viscously dissipated energy can be spent to launch the outflows.
It results in a larger fraction $f=(\dot{M}_0-\dot{M})/\dot{M}_0$ of material launched from the disc (see, e.g., \citealt{1973A&A....24..337S} and simulations by \citealt{2018ApJ...853...45T}). 
The ratio of the total momentum carried by the wind to the total momentum of X-ray photons emitted at the NS surface
\beq 
\frac{\Upsilon_{\rm wind}}{\Upsilon_{X,0}}\approx 
\frac{8}{5}\frac{f}{1-f}
\eeq 
exceeds unity at $f>5/13$ already.
However, in the case of strong geometrical beaming, X-ray photons experience multiple reflections from the walls of the accretion cavity and can transfer their momentum to the outflow several times. 
Thus, the effective momentum transferred to the outflow by X-ray photons $\Upsilon_X$ (depending on the geometry of the cavity) can be larger than the total momentum of photons emitted at the NS surface, i.e., $\Upsilon_X/\Upsilon_{X,0}>1$ (see Fig.\,\ref{pic:sc_Lp}).
Thus, geometrical beaming is associated with increased radiative pressure on the walls of the accretion cavity. 
Under conditions of sufficiently strong beaming, the momentum transfer from X-rays to the outflow becomes comparable to the total momentum of the outflow ($\Upsilon_X/\Upsilon_{\rm wind}\approx 1$, see horizontal lines in Fig.\,\ref{pic:sc_Lp}) even in the case of a large fraction of material launched with the wind. 
It might increase the opening angle of the accretion cavity and the corresponding reduction of the geometrical beaming. 
As a result, we expect that the fractional mass outflow rate puts an upper limit on the opening angle of the accretion cavity and possible amplification factor.
Prominent amplification factors require a large fractional mass outflow rate.
Detailed analysis requires \red{radiative hydrodynamics} simulations, which is beyond this paper's scope.

%---------------------------------------------%
%-----------------------------------------------%

%%%%%%%%%%%%%%%%%%%%%%%%%%%%%%%%%%%%%%%%%%%%%%%%%
\section{Summary and discussion}
\label{sec:Sum}
%%%%%%%%%%%%%%%%%%%%%%%%%%%%%%%%%%%%%%%%%%%%%%%%%

Using Monte-Carlo simulations, we have investigated the influence of the outflows in bright XRPs on the geometrical beaming of radiation and key features of pulsations.
We have considered a more realistic geometry of the accretion cavity (see Fig.\,\ref{pic:scheme}) compared to those investigated earlier in \citealt{2021MNRAS.501.2424M}, and we have the influence of outflow velocity taken into account. 
We confirm that strong geometrical beaming is inconsistent with a large PF: the amplification of the apparent luminosity by order of magnitude (or more) results in a drop in a PF of less than 10 per cent.
This weak response is essential for understanding the physics of pulsating ULXs, and it suggests that the actual accretion luminosity in these objects cannot differ much from their apparent luminosity.
Our conclusion does not rule out the possibility of outflows to ULXs. 
On the contrary, given the estimates (see \ref{eq:dotM_lim_d} and \ref{eq:Mdot_lim_q}) of the accretion rate needed to initiate the outflows mass losses due to the radiation-driven winds in ULXs are likely.
A change in the outflow geometry (its opening angle or the orientation with respect to the observer's reference frame) or the on/off-switching of outflows can initiate changes in the ULX's apparent luminosity.

We demonstrated that geometrical beaming causes phase shifts in the pulsations of the X-ray energy band.
Therefore, detecting lags in the pulsations in transient XRPs while they are transiting to a bright state seems to correspond to the launch of the outflows from the accretion disk. 
Phase lags arising at high mass-accretion rates were reported in a few X-ray transients such as
in Swift~J0243.6+6124 at $L\sim 10^{38}\,\ergs$ \citep{2020MNRAS.491.1857D}, in RX~J0209.6-7427 at $L\sim 2\times 10^{38}\,\ergs$ \citep{2022arXiv220814785H}, in SMC~X-3 at $L\sim 2\times 10^{38}\,\ergs$ \citep{2022arXiv220911496L}, and in GRO~J1744-28 during the short bursts on top of the outburst, when the apparent luminosity of the source increases by a factor of $10$ or more and exceeds $10^{39}\,\ergs$ \citep{1996Natur.379..799K,2015MNRAS.452.2490D}.
When accounting for the limiting mass accretion rate required to launch the radiation-driven outflow (see estimations \ref{eq:dotM_lim_d} and \ref{eq:Mdot_lim_q}), we conclude that the outflows and corresponding phase lags are expected at apparent luminosity $L>10^{39}\,\ergs$. 
As a result, the phase shifts detected in GRO~J1744-28 can already be considered as a possible consequence of the launching of the outflow.
The phase shifts and changes of the pulse profiles in Swift~J0243.6+6124, RX~J0209.6-7427, and SMC~X-3 can be caused by changes in the geometry of the emitting region near the NS's surface \red{(i.e., the appearance of accretion column \citealt{1976MNRAS.175..395B,1981A&A....93..255W,2015MNRAS.447.1847M,2022arXiv220712312A,2022MNRAS.515.4371Z,2022arXiv221006616Z}).} These then correspond to changes in the beam pattern, see, e.g., \citealt{1973A&A....25..233G} or in the growth of the accretion column up to the height where it can not be eclipsed by the NS \citealt{2018MNRAS.474.5425M}). 
Within the opening angle of the accretion cavity formed by the outflows, phase lags tend to be more sensitive to the observer's viewing angle than to the luminosity amplification factor and the PF (see Fig.\,\ref{pic:amp_and_PF}).
Variations in outflow geometry, and the corresponding changes of the phase lags, can influence the detectability of pulsation in bright XRPs and in ULXs powered by accretion onto the NS.
Precession of the outflow and the accretion cavity in the observer's frame of reference may influence the apparent pulse period derivative.

Our simulations account for the effect of special relativity and possibly the large velocity of the outflow. Taking these effects into account is important even though the outflow velocity $\lesssim 0.2c$ according to the recent observational results \citep{2018MNRAS.479.3978K,2019MNRAS.487.4355V,2021MNRAS.505.5058P}.
Accounting for the high velocity of the outflows, however, hardly affects the luminosity amplification through geometrical beaming (see Fig.\,\ref{pic:sc_beaming_velosity}a).
The influence of the velocity on the PF is also negligible (see Fig.\,\ref{pic:sc_beaming_velosity}b). 

We demonstrated that strong geometrical beaming is associated with large radiation pressure on the walls of the accretion cavity because photons experience multiple reflections from the cavity walls.
In the case of weakly advective accretion discs, 
a considerable (if not the major) fraction of the material accreting from the companion star is lost in the wind.
Then geometry of the cavity is not affected by radiation pressure.
On the contrary, in the case of strong advection, discs lose only a small fraction of material due to the winds. 
Then high radiation pressure can affect the opening angle of the accretion cavity, making it larger, which reduces the geometrical beaming.

%%%%%%%%%%%%%%%%%%%%%%%%%%%%%%
\section*{Acknowledgements}
%%%%%%%%%%%%%%%%%%%%%%%%%%%%%%

This work was supported by UKRI Stephen Hawking fellowship and the Netherlands Organization for Scientific Research Veni Fellowship (AAM).
The authors thank Matthew Middleton, Juri Poutanen, and Sergey Tsygankov for critical remarks and motivating discussion.

%%%%%%%%%%%%%%%%%%%%%%%%%%%%%%%%%
\section*{Data availability}
%%%%%%%%%%%%%%%%%%%%%%%%%%%%%%%%%

The calculations presented in this paper were performed using a private code developed and owned by the corresponding author. All the data appearing in the figures are available upon request. 

%%%%%%%%%%%%%%%%%%%%%%%%%%%%%%%%%%%%%%%%%%%%%%%%%%%%%%%%%%%%%%%%%%%%%%%%%%%%%%
%% Bibliography %%
%%%%%%%%%%%%%%%%%%%%%%%%%%%%%%%%%%%%%%%%%%%%%%%%%%%%%%%%%%%%%%%%%%%%%%%%%%%%%%
%\bibliographystyle{mn2e}
%\bibliographystyle{mnras}
%\bibliography{allbib}

\appendix

%\section{ToDo}
%\label{App:Code}

%\red{
% False spin period derivatives. Estimation of the spin period derivative. 
%}

%Phase lags are small for the strongest amplification. Do we see J1744 face on? \\
%Variability of ULXs due to the appearance of the outflows? As soon as outflows arise, the luminosity of the source drops. \\
%Note that only ULXs can be beamed in the case of normal magnetic field strength. \\
%Phase shift varying with time and detectability of pulsation in ULXs. \\
%Outflow should arise above $10^{40}\,\ergs$. How do they affect the luminosity distribution function? Is it possible that the cut-off is due to the beaming?

%- pulse topology
%- influence of absorption ?

% Don't change these lines
\bsp % typesetting comment
\label{lastpage}
\end{document}